\documentclass[amstex,twocolumn,showpacs,floats,floatfix,superscriptaddress,aps,pra]{revtex4}
\usepackage{amsfonts}
\usepackage{amssymb}
\usepackage{amsmath}
\usepackage{calc}
\usepackage{graphicx}
\usepackage{bm}

\def\be{ \begin{equation} }
\def\ee{ \end{equation} }
\def\bea{ \begin{eqnarray} }
\def\eea{ \end{eqnarray} }
\def\ba{ \begin{array} }
\def\ea{ \end{array} }
\def\bse{\begin{subequations}}
\def\ese{\end{subequations}}
\def\half{\tfrac12}
\def\state#1{\left\vert \psi _{#1}\right\rangle}
\def\state#1{ \psi _{#1}}
\def\adbstate{\varphi}

\def\ee{ \end{equation} }

\begin{document}

\author{A. A. Rangelov}
\affiliation{Department of Physics, Sofia University, James Bourchier 5 blvd, 1164 Sofia, Bulgaria}
\affiliation{Unit\`a CNISM, Dipartimento di Fisica E. Fermi, Universit\`a di Pisa, Largo Pontecorvo 3, I-56127 Pisa, Italy}
\author{N. V. Vitanov}
\affiliation{Department of Physics, Sofia University, James Bourchier 5 blvd, 1164 Sofia, Bulgaria}
\affiliation{Institute of Solid State Physics, Bulgarian Academy of Sciences, Tsarigradsko chauss\'{e}e 72, 1784 Sofia, Bulgaria}
\author{E. Arimondo}
\affiliation{Unit\`a CNISM, Dipartimento di Fisica E. Fermi, Universit\`a di Pisa, Largo Pontecorvo 3, I-56127 Pisa, Italy}
\title{Stimulated Raman adiabatic passage into continuum}
\date{\today }

\begin{abstract}
We propose a technique, which produces nearly complete ionization of the population of a discrete state,
 coupled to a continuum by a two-photon transition via a lossy intermediate state, whose lifetime is much shorter than the interaction duration. 
We show that using counterintuitively ordered pulses, as in stimulated Raman adiabatic passage (STIRAP),
 wherein the pulse coupling the intermediate state to the continuum precedes and partly overlaps the pulse coupling the initial and intermediate states,
 greatly increases the ionization signal and strongly reduces the population loss due to spontaneous emission through the lossy state. 
For strong spontaneous emission from that state, however, the ionization is never complete because the dark state required for STIRAP does not exist. 
We demonstrate that this drawback can be eliminated almost completely by creating a laser-induced continuum structure (LICS)
 by embedding a third discrete state into the continuum with a third, control laser. 
This LICS introduces some coherence into the continuum, which enables a STIRAP-like population transfer into the continuum.
A highly accurate analytic description is developed and numerical results are presented for Gaussian pulse shapes.
\end{abstract}
\maketitle


\section{Introduction}


Stimulated Raman adiabatic passage (STIRAP) is a simple, robust and efficient technique for complete population transfer (CPT) in three-state quantum systems \cite{STIRAP}. 
In this technique, the population is transferred adiabatically in a Raman transition, from an initially populated state $\state{1}$ via an intermediate state $\state{2}$
 to a target state $\state{3}$ by two pulsed fields, pump and Stokes, whose frequencies are maintained on two-photon resonance between states $\state{1}$ and $\state{3}$. 
If the pulses are ordered counterintuitively, the Stokes before the pump, then the dark state is associated with state $\state{1}$ initially and state $\state{3}$ in the end,
 thus providing an adiabatic route from $\state{1}$ to $\state{3}$. 
A unique and remarkable feature of STIRAP is that during the transfer the population remains trapped in a dark state,
 which is a time-dependent coherent superposition of states $\state{1}$ and $\state{3}$ only and does not involve the intermediate state $\state{2}$. 
State $\state{2}$ therefore remains unpopulated during the transfer and its properties, including possible population decay, are largely irrelevant for STIRAP.
A very large detuning \cite{STIRAP-detuning} or loss rate \cite{STIRAP-loss}, however, do affect STIRAP and reduce its transfer efficiency.

The simplicity, efficiency, and robustness of STIRAP have attracted much attention, which has resulted in numerous applications in a variety of quantum systems. 
Among them, we mention population transfer via a continuum, wherein the discrete intermediate state $\state{2}$ is replaced by an ionization continuum \cite{STIRAP-continuum,hydrogen,Peters}. 
Because a dark state that links adiabatically states $\state{1}$ and $\state{3}$ still forms, high transfer efficiency is still possible.
However, various specific features of the continuum that make it substantially different from a discrete state (or a manifold of discrete states) reduce the transfer efficiency. 
Several scenarios have been proposed to reduce the effects of these continuum features, resulting in increase of the transfer efficiency well above 50\%. 
STIRAP via continuum has recently been demonstrated experimentally \cite{Peters}. 

The extension of STIRAP to include an initial or final continuum was at first investigated for the photoassociation process producing cold molecules starting from laser-cooled atoms \cite{Javanainen98}. 
Photoassociation is based on the laser-driven transitions of a continuum-bound-bound system. 
Vardi \emph{et al} \cite{Vardi99} have extended the STIRAP technique to the cases of initial or final continuum, proving that transfer with a good efficiency is possible.
The quantum transfer in a three-level system based on bound-bound-continuum transitions arises also in the ionization of a Bose-Einstein condensate (BEC). 
Ionization of a rubidium BEC from the ground state through a two-photon ionization (TPI) scheme was explored experimentally \cite{Ciampini02} and theoretically \cite{Anderlini06}.

In the present paper, we follow this latter idea and introduce a continuum in STIRAP by replacing the final state $\state{3}$ by a continuum. 
The objective, compared to STIRAP \emph{via} a continuum, is also changed drastically: instead of trying to \emph{avoid} ionization, here we aim at \emph{maximizing} ionization. 
As in STIRAP we also aim at minimizing the transient population of the intermediate state $\state{2}$,
 in order to avoid (possibly strong) decay to other discrete states via spontaneous emission or unwanted excitation to other states. 
States $\state{1}$ and $\state{2}$ are linked by a pump pulse, and state $\state{2}$ is connected to the continuum by an ionizing pulse, with the ionizing pulse arriving before the pump. 
The challenge here is that, unlike STIRAP \emph{via} continuum, a dark state cannot be formed between the initial state $\state{1}$ and the continuum states, because a flat continuum is an incoherent medium.

To this end, we propose to use a laser-induced continuum structure (LICS) \cite{LICS} created in the continuum
 by embedding an ancillary, control state $\state{c}$ into the continuum by a third, control laser. 
This LICS creates some coherence into the continuum, which makes it possible to create a \emph{quasi-dark} state,
 and thence a STIRAP-like process into the continuum. 
This LICS-STIRAP allows us to produce almost complete ionization, with negligibly small population losses from state $\state{2}$,
 even when the $\state{2}$ lifetime is much shorter than the interaction duration.  
Experimental verification of this LICS-STIRAP technique will open opportunities for many applications,
 for instance efficient photoionization of a BEC without atomic excitation in intermediate atomic states that will improve the ion production obtained in the experiment of ref. \cite{Ciampini02}.
In addition, an increase in the ionization efficiency for cold atoms will improve the brightness in magneto-optical-trap based sources producing either electron \cite{Claessens05} or ion \cite{Hanssen06} beams. 
 
The structure of this paper is as follows. 
In Sec. \ref{Sec:c-STIRAP} we define the problem and show that a counterintuitive pulse order of the pump and ionizing pulses
 suppresses fluorescence and is favourable for ionization, even without a control laser.
In Sec. \ref{Sec:LICS-STIRAP} we add LICS to the scenario and derive a very accurate analytic approximation to describe this LICS-STIRAP technique.
Section \ref{Sec:Examples} provides illustrations of the proposed ionization technique and comparison of the analytical and numerical results. 
Section \ref{Sec:Conclusions} presents a summary.


\section{STIRAP into continuum\label{Sec:c-STIRAP}}

\begin{figure}[tb]
\centering\includegraphics[height=80mm]{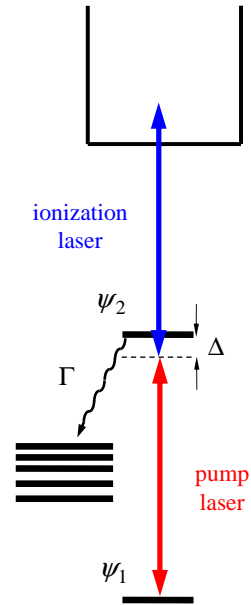}
\caption{(Color online) TPI scheme. An initially populated discrete state $\state{1}$ is coupled to another discrete state $\state{2}$ by a resonant, or nearly resonant, pump laser field.
State $\state{2}$ is coupled to the ionization continuum by a second, ionizing laser pulse.
State $\state{2}$ can decay irreversibly  to other states.
}
\label{Fig:c-STIRAP}
\end{figure}

We first consider TPI  of an atom, initially in state $\state{1}$, coupled to the ionization continuum via state $\state{2}$, as illustrated in Fig. \ref{Fig:c-STIRAP}. 
The transition 1-2 is driven by a pump laser pulse with Rabi frequency $\Omega (t)$ and detuning $\Delta$,
 and state $\state{2}$ is connected to the continuum by a second laser pulse with a time-dependent rate $\Gamma_i(t)$. 
State $\state{2}$ can decay irreversibly via spontaneous emission (or other mechanisms) to other states with a constant rate $\Gamma$;
 we shall refer to the respective signal as the \emph{fluorescence signal},
\be
F=\int_{-\infty }^{\infty }\Gamma P_2(t)dt.  \label{fluorescence}
\ee
The ionization signal is 
\be
I=1-P_1(\infty )-P_2(\infty )-F, \label{ionization}
\ee
where $P_n(t)$ are the populations of the discrete states ($n=1,2$).
Our objective is to set up the laser pulses such that the ionization signal $I$ is maximized, while the fluorescence signal $F$ is minimal.

We wish to design a recipe to maximize ionization when the loss rate $\Gamma$ is large compared to the interaction duration $T$. In other words, we wish to ionize the atom, without exciting it,
 despite being on resonance with state $\state{2}$. 
This objective reminds one of STIRAP. The significant difference here is that state $\state{3}$ is replaced by a continuum. 

The equation that describes the dynamics of the system (in units $\hbar =1$) is the Schr\"odinger equation, 
\be
i\frac{d}{dt}\mathbf{c}(t) = \mathsf{H}(t) \mathbf{c}(t).  \label{Schroedinger}
\ee
Here $\mathbf{c}(t)=\left[ c_1(t),c_2(t)\right]^{T}$ is the column-vector with the probability amplitudes $c_1(t)$ and $c_2(t)$ of states $\state{1}$ and $\state{2} $,
 and $\mathsf{H}(t)$ is the Hamiltonian, obtained by adiabatic elimination of the continuum states and within the rotating-wave approximation (RWA) \cite{LICS},
\be
\mathsf{H}(t) = \tfrac12\left[\ba{cc}
2S_1(t) & \Omega (t) \\ 
\Omega (t) & 2\Delta + 2S_2(t) -i\Gamma_i(t) -i\Gamma
\ea\right],
\ee
where $S_1(t)$ and $S_2(t)$ are the Stark shifts of states $\state{1}$ and $\state{2}$, produced by virtual excitation to other atomic states.

We are interested in situations when the ionization signal $I$ is large, i.e. in the non-perturbative regime. 
This implies large peak Rabi frequency $\Omega (t)$ and large ionization rate $\Gamma_{i}(t)$.
Because we also assume that the loss rate $\Gamma $ is fixed and large, $\Gamma \gg 1/T$,
 state $\state{2}$ is subjected to strong population decay, due to both spontaneous emission and ionization. 
This implies that it receives very little transient population and can therefore be eliminated adiabatically. 
Hence we find after simple algebra
\bse\label{P1P2-ae}\bea
P_1(t) &\approx & \exp \left[ - \int_{-\infty }^{t}\frac{\Omega(t^\prime )^2 \left[ \Gamma_i (t^\prime )+\Gamma \right] }
 {\left[\Gamma_i(t^\prime)+\Gamma \right]^2 +4\left[ \Delta +S(t^\prime )\right]^2}dt^\prime \right] ,  \label{P1(t)} \\
P_2(t) &\approx & \frac{\Omega (t)^{2}}{\left[ \Gamma_i(t)+\Gamma \right] ^2 +4\left[ \Delta + S(t)\right]^2} P_1(t), \label{P2(t)}
\eea\ese
with $S(t)=S_2(t)-S_1(t)$.
Now we use these formulas to examine the possibilities of how to minimize fluorescence $F$, i.e. $P_2(t)$, and simultaneously maximize ionization $I$.

The first choice is to use a \emph{large detuning} $\Delta$. 
Indeed, this will reduce the population of state $\state{2}$, because $P_2(t)\sim \Delta^{-2}$ for large $\Delta$ [see Eq. \eqref{P2(t)}];
 however, this decrease will be accompanied by an increase in the population of state $\state{1}$, see Eq. \eqref{P1(t)}. 
In result, the increase in the ionization will be little, if any.
A similar conclusion applies to the Stark shift $S (t)$, which can be induced by the ionizing laser,
 or, if needed, by an additional far-off-resonance laser, as in SCRAP technique \cite{SCRAP}.

A straightforward alternative is to increase the magnitude of the ionizing pulse $\Gamma _{i}(t)$ alone. 
Then the population (\ref{P2(t)}) of state $\state{2}$ decreases as $\Gamma _{i}^{-2}$, as vs $\Delta $;
 however, the increase in the population (\ref{P1(t)}) of state $\state{1}$ is smaller vs $\Gamma _{i}$ than vs $\Delta $. 
In result, the ionization signal will increase more markedly when increasing $\Gamma _{i}$.

\begin{figure}[tb]
\centering\includegraphics[width=65mm]{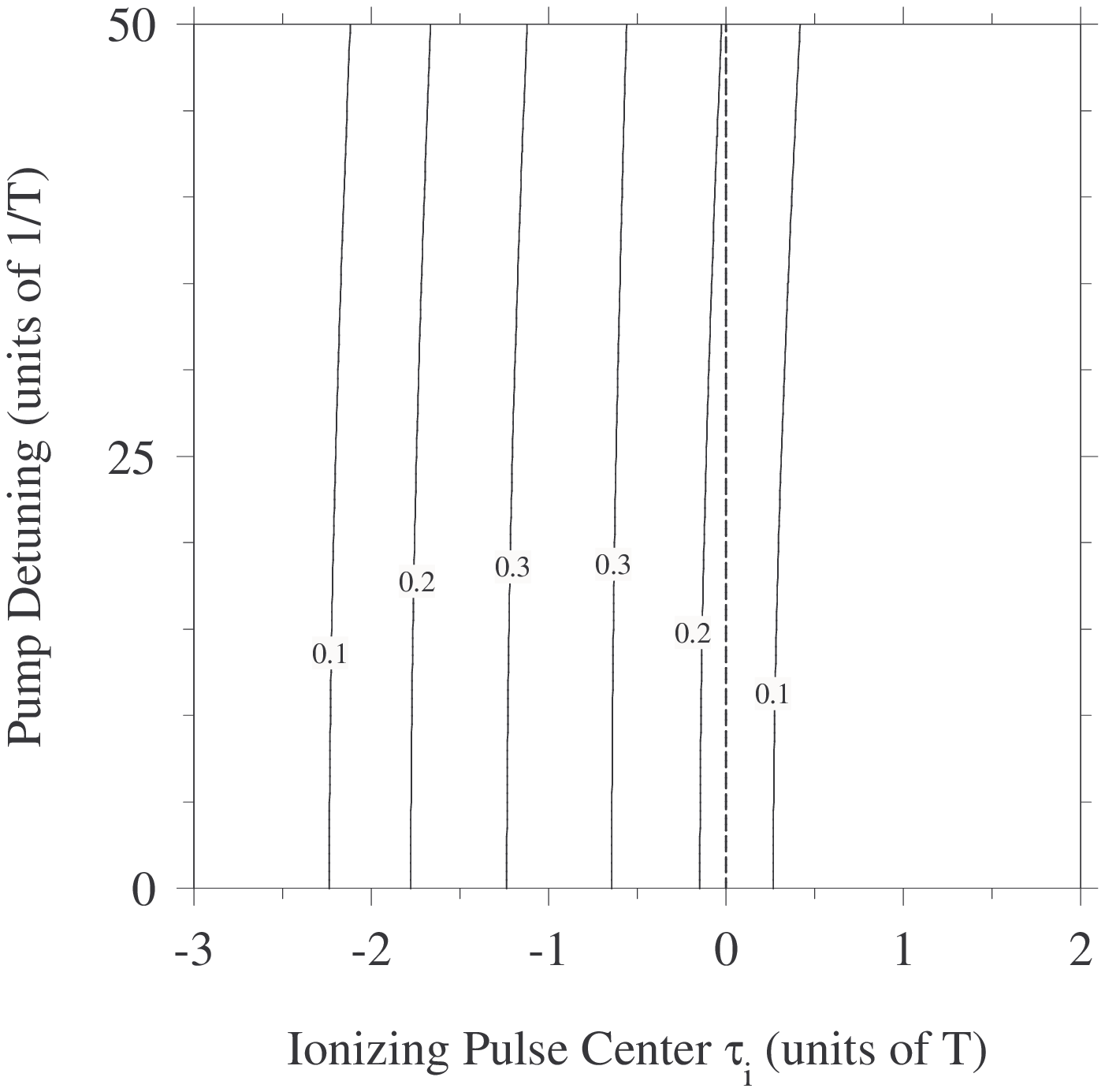}
\caption{Contour plots of the ionization signal $I$ as a function of the delay of the ionizing pulse $\protect\tau _{i}$ and the pump detuning $\Delta =0$. 
The pulses have Gaussian shapes, Eqs. \eqref{Gaussians}.
The decay rate from state $\state{2}$ is $\Gamma =100/T$, the peak ionization rate is $\Gamma_{i0}=50/T$, the peak Rabi frequency is $\Omega_0 =50/T$, and the Stark shift is $S=0$.}
\label{Fig:td}
\end{figure}

A closer inspection of Eqs. \eqref{P1P2-ae} suggests that one can decrease $P_{2}(t)$, without increasing $P_{1}(t)$
 (implying thence a net increase of the ionization $I$), by delaying the pump pulse $\Omega(t) $ with respect to the ionizing pulse $\Gamma _{i}(t)$. 
Indeed, it is obvious that the pump pulse $\Omega (t)$ must not arrive \emph{before} the ionizing pulse $\Gamma (t)$,
 because then the fluorescence will deplete the population even before ionization has the chance to begin;
 mathematically, this implies large values for the fractions in Eqs. \eqref{P1P2-ae}, with resulting small population of state $\state{1}$ and large fluorescence signal $F$. 
In contrast, if the pump pulse $\Omega (t)$ arrives simultaneously, or \emph{after} the ionizing pulse $\Gamma _{i}(t)$, with some overlap,
 then fluorescence can only begin simultaneously with ionization. 
Moreover, if during the ionization the ratio $\Omega (t)/\Gamma _{i}(t)$ is very small,
 while the ratio $\Omega(t)^{2}/\Gamma _{i}(t)$ is moderately large compared to $1/T$,
 then both $P_{1}(t)$ and $P_{2}(t)$ will remain small, as easily seen from Eqs. \eqref{P1P2-ae} when $\Delta =S =0$. 
Therefore, our objective of producing ionization without excitation requires that during the population depletion we have 
\be
\Omega (t)^2 T \gtrsim \left[ \Gamma_i(t)+\Gamma \right] \gg \Omega (t). \label{c-STIRAP conditions}
\ee
These conditions require large pulse areas over the interaction duration,
\be
\int_{-\infty}^\infty \Omega (t)dt \gg 1, \quad \int_{-\infty}^\infty \Gamma_{i} (t)dt \gg 1.
\ee
Conditions (\ref{c-STIRAP conditions}) suggest that the pump pulse Rabi frequency $\Omega (t)$ should be small in comparison with the ionizing rate $\Gamma_i(t)$. 
This can be naturally achieved, indeed, if the pump pulse is delayed to, but overlapped with, the ionizing pulse. 
Then ionization will occur during the rising edge of the pump pulse.

\begin{figure}[tb]
\centering\includegraphics[width=65mm]{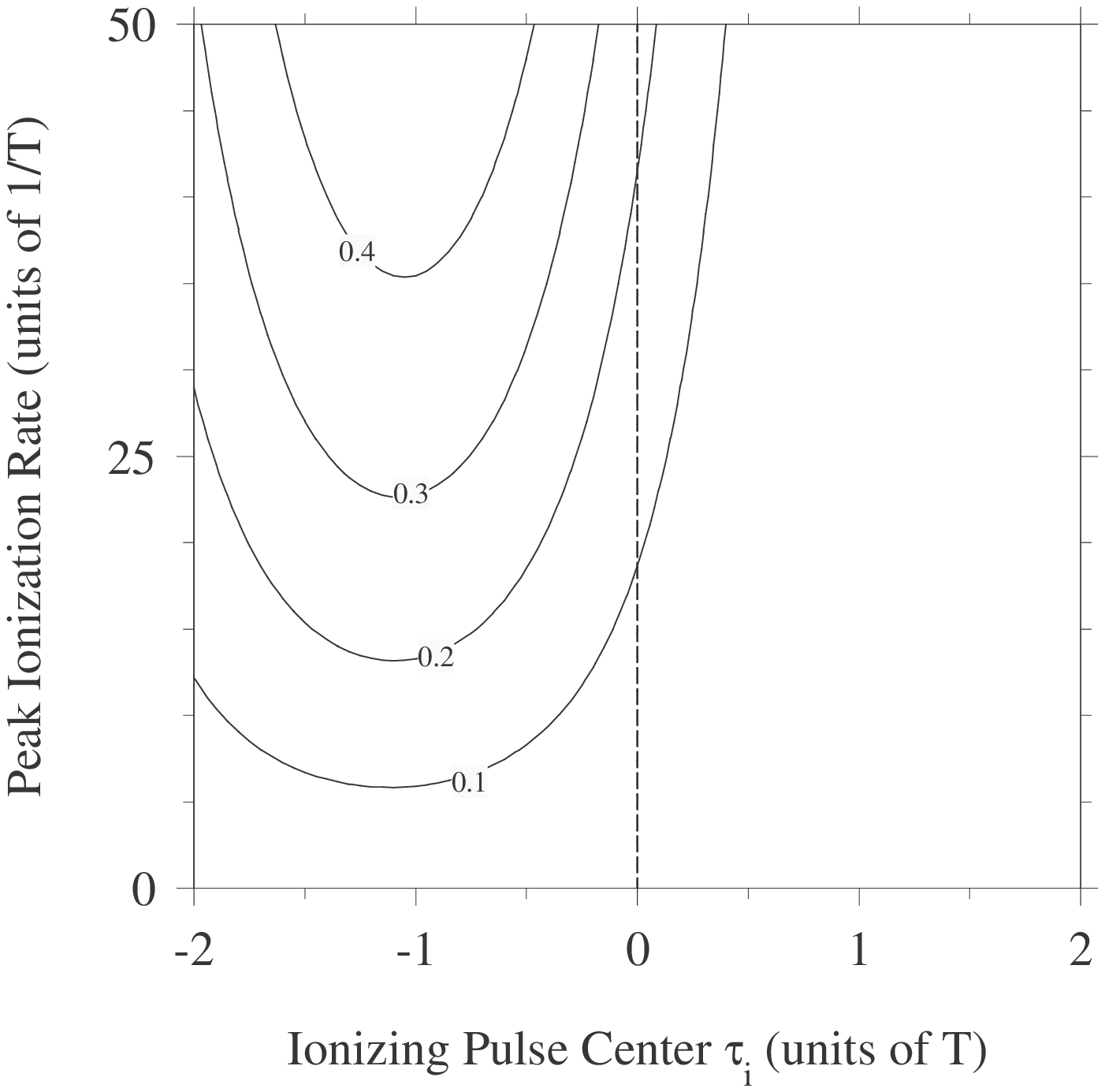}
\caption{Contour plots of the ionization signal $I$ as a function of the delay of the ionizing pulse $\tau_i$ and the peak ionization rate $\Gamma_{i0}$. 
The pulses have Gaussian shapes, Eqs. \eqref{Gaussians}.
The pump detuning is $\Delta =0$, the decay rate from state $\state{2}$ is $\Gamma=100/T$, the peak Rabi frequency is $\Omega_0 =50/T$, and the Stark shift is $S=0$.}
\label{Fig:ts}
\end{figure}

These conclusions are illustrated in Figs. \ref{Fig:td} and \ref{Fig:ts} where the ionization signal is plotted as a function of, respectively,
 the ionizing pulse delay and the pump pulse detuning $\Delta$, and the ionizing pulse delay and the ionizing pulse intensity. 
These figures clearly demonstrate that the counterintuitive pulse order -- ionizing pulse before pump pulse -- is favorable for ionization.
Figure \ref{Fig:td} demonstrates also that the detuning $\Delta$ is of little help in respect to ionization, as predicted by Eqs. \eqref{P1P2-ae}. 

As follows from the above analysis, counterintuitive pulse order increases ionization and suppresses excitation. 
However, very strong ionizing laser is needed to ensure the conditions \eqref{c-STIRAP conditions}. 
In the following section we shall show that a LICS in the continuum can help ionization and make this process very similar to STIRAP
 because of the creation of a quasi-dark state between state $\state{1}$ and the LICS, with ensuing nearly complete ionization with moderate laser resources.


\section{STIRAP into LICS\label{Sec:LICS-STIRAP}}


\subsection{The system}


\begin{figure}[tb]
\centering\includegraphics[height=80mm]{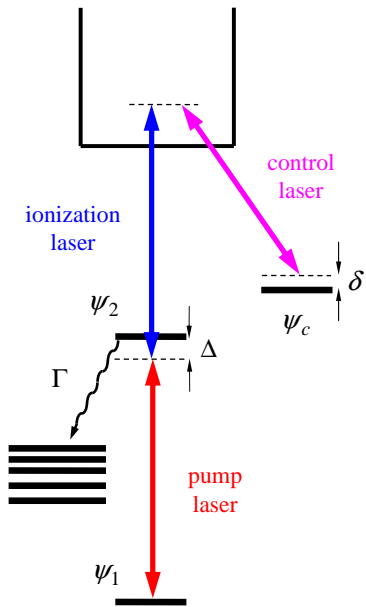}
\caption{(Color online) The scheme for ionization by LICS-STIRAP, which extends the scheme in Fig. \ref{Fig:c-STIRAP}. 
The additional discrete state $\state{c}$ is embedded into the continuum by a control laser field, which creates a LICS in the continuum.}
\label{Fig:LICS-STIRAP}
\end{figure}

Let us assume now that, in addition to the scheme in Fig. \ref{Fig:c-STIRAP},
 an additional laser pulse couples a third discrete state $\state{c}$ with the continuum, as shown in Fig. \ref{Fig:LICS-STIRAP}. 
The dynamics of the system is again described by Eq. \eqref{Schroedinger},
 with $\mathbf{c}(t)=\left[ c_1(t), c_2(t),c_c(t)\right] ^{T}$.
The Hamiltonian after adiabatic elimination of the continuum states and within the RWA reads
\be
\mathsf{H}=\half\left[ \begin{array}{ccc}
2S_1 & \Omega & 0 \\ 
\Omega & 2\Delta +2S_2 - i\Gamma_i - i\Gamma & -(q+i)\sqrt{\Gamma_i\Gamma_c} \\ 
0 & -(q+i)\sqrt{\Gamma_i\Gamma_c} & 2\delta + 2S_c - i\Gamma_c
\end{array}\right] , \label{Hamiltonian}
\ee
where explicit time dependences are omitted for brevity.

The constant $q$, called the Fano parameter \cite{Fano}, is an important feature of LICS.
It is responsible for the asymmetric dependence of the ionization signal on the two-photon detuning \cite{LICS}, and 
 it also plays an important role in population transfer via continuum \cite{STIRAP-continuum}. 
The quantities $\Gamma_2=\sum_\alpha \Gamma_2^\alpha$ and $\Gamma_c=\sum_\alpha \Gamma_c^\alpha$ ($\alpha=p,i,c$)
 are the total ionization rates of states $\state{2} $ and $\state{c}$, respectively,
 which are given by sums of ionization rates induced by the pump (p), ionization (i) and control (c) pulses,
 whereas $S_n=\sum_\alpha S_n^\alpha$ ($n=1,2,c$) are the dynamic Stark shifts. 
The ionization widths and the Stark shifts are proportional to the pulse intensities $I_p(t)$, $I_i(t)$ and $I_c(t)$,
 $\Gamma_n^\alpha (t) = \Gamma_{n0}^\alpha I_\alpha (t)$ and $S_n^\alpha(t)=S_{n0}^\alpha I_\alpha(t)$,
 where the parameters $\Gamma_{n0}^\alpha$ and $S_{n0}^\alpha$ depend on the particular atomic states and the laser frequencies.
 
For simplicity, we will assume that the ionization of state $\state{2}$ occurs due to the action of the ionizing laser $I_i$ only: $\Gamma_2=\Gamma_2^i$;
 this condition can be satisfied by selecting appropriate atomic levels and laser frequencies
 such that the pump and control lasers do not cause direct ionization from state $\state{2}$.
We will also assume that the ionization rate of state $\state{c}$ is induced by the control laser $I_c$ only: $\Gamma_c=\Gamma_c^c$.
Given the preceding assumption for $\Gamma_2$ it is clear that the ionizing laser $I_i$ will ionize also state $\state{c}$ ($\Gamma_c^i\neq 0$).
However, because from state $\state{c}$ the ionizing laser points deeply into the continuum this ionization rate is small;
 moreover, as state $\state{c}$ remains largely unpopulated, this ionization channel (which is actually favorable for our goal of maximizing ionization)
 does not alter the dynamics markedly.

In addition, we also neglect the Stark shifts, $S_1=S_2=S_c=0$; 
 these are important in population transfer via continuum \cite{STIRAP-continuum}, but do not have much effect here.

The fluorescence signal $F$ from state $\state{2} $ is given by Eq. \eqref{fluorescence},
 and the ionization signal by
\be
I=1-P_{1}(\infty )-P_{2}(\infty )-P_{c}(\infty )-F. \label{ionization3}
\ee

The optimal pulse order of the three pulsed fields is determined by the objective to maximize ionization.
We have already come to the conclusion that the pump and ionizing pulses must arrive in a counterintuitive order: ionizing before pump.
The optimal timing of the control pulse can be deduced from the following arguments.

Because the objective is ionization, we must avoid population transfer via the continuum into state $\state{c}$, 
 which will occur if the control pulse precedes the ionizing pulse \cite{STIRAP-continuum}; the control should therefore be applied after the ionizing pulse.
The timing of the control pulse with respect to the pump pulse is not so significant but these pulses should not be separated too much
 because it is obvious that, for LICS to have any effect, the control pulse must overlap significantly with the pump and ionizing pulses.
We therefore conclude that for maximal ionization, the pulses should be applied in the order \emph{ionizing-control-pump},
 with a sufficient overlap between them.
We shall therefore assume this pulse ordering in the analytical description in the next section,
 which will be confirmed as optimal also by numerical simulations in Sec. \ref{Sec:Examples}.


\subsection{Analytical description}


It is appropriate to describe the evolution of the system in the basis of the instantaneous eigenstates of the Hamiltonian \eqref{Hamiltonian}.
Because this Hamiltonian is non-Hermitian its eigenvalues $\varepsilon_{\alpha }(t)$ are complex valued,
 and the right eigenvectors $\adbstate_\alpha(t)$ differ from Hermitian conjugates of the left eigenvectors  \cite{steering}. 
The right eigenvectors, and the eigenvalues, are defined by the equation ($\alpha =+,0,-$)
\be
\mathsf{H}(t)\adbstate _{\alpha }(t)=\varepsilon _{\alpha }(t)\adbstate _{\alpha}(t),  
\label{right}
\ee
where $\adbstate _{\alpha }(t)$ $=\left[ f_{1\alpha }(t),f_{2\alpha}(t),f_{3\alpha }(t)\right] ^{T}$.
Using such states we expand the state vector as 
\be
\mathbf{\Psi }(t)=b_{+}(t)\adbstate _{+}(t)+b_{0}(t)\adbstate _{0}(t)+b_{-}(t)\adbstate_{-}(t).
\ee
The probability amplitudes in the original basis and the adiabatic basis are related through the transformation 
\be
\mathbf{c}(t)=\mathsf{R}(t)\mathbf{b}(t),  \label{bare_states}
\ee
where the column vector $\mathbf{b}(t)=\left[ b_{+}(t),b_{0}(t),b_{-}(t)\right]^{T}$ comprises the probability amplitudes of the adiabatic states.
The columns of $\mathsf{R}$ are the components of $\adbstate_\alpha(t)$, $\left(\mathsf{R}(t)\right)_{n\alpha}=f_{n\alpha}(t)$,
 with $n=1,2,3$ and $\alpha =+,0,-$. 

The Schr\"{o}dinger equation for the vector $\mathbf{b}$ reads%
\be
i\frac{d}{dt}\mathbf{b}(t)=\mathsf{H}^{ad}(t)\mathbf{b}(t),
\label{Schroedinger equation for the vector B}
\ee
where $\mathsf{H}^{ad}(t)=\mathsf{H}^{a}(t)+\mathsf{H}^{na}(t)$, with an adiabatic diagonal Hamiltonian 
\be
\mathsf{H}^a(t)=\mathsf{R}^{-1}(t)\mathsf{H}(t)\mathsf{R}(t)=\left[ \begin{array}{ccc}
\varepsilon _{+}(t) & 0 & 0 \\ 
0 & \varepsilon _{0}(t) & 0 \\ 
0 & 0 & \varepsilon _{-}(t)%
\end{array}\right] ,
\ee
and a nonadiabatic coupling $\mathsf{H}^{na}(t)=-i\mathsf{R}^{-1}(t)\frac{d}{dt}\mathsf{R}(t)$.
If the time evolution is slow we can neglect the nonadiabatic coupling;
 then Eq. \eqref{Schroedinger equation for the vector B} is easily solved,
\be
b_{\alpha }(t)=b_{\alpha }(-\infty )\exp \left[ -i\int_{-\infty}^{t}\varepsilon _{\alpha }(t^{\prime })dt^{\prime }\right] \quad (\alpha=+,0,-).
\label{adiab_solution}
\ee

Our major approximation is based on the assumption that the population dynamics takes place mainly during the time interval when $\Omega\ll \sqrt{\Gamma_i\Gamma_c}$.
 i.e. during the rising edge of the pump pulse (which is the last to arrive).
This assumption derives from our interest in the regime of large ionization, which requires strong laser fields.
In this case we can approximate the eigenvalues and the eigenvectors by assuming that the ratio $\Omega/\sqrt{\Gamma_i\Gamma_c}$ is a small parameter \cite{steering}. 
After simple algebra we obtain
\bse\bea
\varepsilon_{+} &\approx&\tfrac12 \left[ \widetilde{\Delta}+\widetilde{\delta} + \eta \right]
 + \frac{\Omega^2}{4\eta }\frac{\eta +\widetilde{\Delta}-\widetilde{\delta}}{\eta +\widetilde{\Delta}+\widetilde{\delta}}, \\
\varepsilon _{0} &\approx&\frac{\Omega^2}{\widetilde{\Gamma}^2-4\widetilde{\Delta}\widetilde{\delta}}\widetilde{\delta}, \\
\varepsilon _{-} &\approx&\tfrac12\left[ \widetilde{\Delta}+\widetilde{\delta}-\eta \right]
 -\frac{\Omega^2}{4\eta }\frac{\eta -\widetilde{\Delta}+\widetilde{\delta}}{\eta -\widetilde{\Delta}-\widetilde{\delta}},
\eea\ese
where
\bse
\bea
\widetilde{\Delta} &=& \Delta -i\frac{\Gamma_i + \Gamma}{2}, \\
\widetilde{\delta} &=& \delta -\frac{i\Gamma_c}{2}, \\
\widetilde{\Gamma} &=& -\left( q+i\right) \sqrt{\Gamma_i\Gamma_c}, \\
\eta &=& \sqrt{\left[ \Delta -\delta -\frac{i\left(\Gamma_i+\Gamma-\Gamma_c\right)}{2}\right] ^2 +\left( q+i\right) ^2 \Gamma_i\Gamma_c}.
\eea
\ese
The corresponding eigenvectors are
\bse\bea
\adbstate _{+}(t) &\approx&\left[ \frac{\Omega}{\widetilde{\Gamma}}\frac{\eta +\widetilde{\Delta}
 -\widetilde{\delta}}{\eta +\widetilde{\Delta}+\widetilde{\delta}}\sin \xi ,\cos \xi,\sin \xi \right]^{T}, \\
\adbstate _{0}(t) &\approx&\left[ 1 - \frac{\Omega^2(\widetilde{\Gamma} ^{2}+4\widetilde{\delta}^{2})}{2(\widetilde{\Gamma} ^{2}-4\widetilde{\Delta}\widetilde{\delta})^{2}},\frac{2\widetilde{\delta}\Omega}{\widetilde{\Gamma} ^{2}-4\widetilde{\Delta}\widetilde{\delta}},-\frac{\Omega
\widetilde{\Gamma}}{\widetilde{\Gamma} ^{2}-4\widetilde{\Delta}\widetilde{\delta}}\right] ^{T},
\label{quasi dark state} \\
\adbstate _{-}(t) &\approx&\left[ \frac{\Omega}{\widetilde{\Gamma} }\frac{\eta -\widetilde{\Delta}+%
\widetilde{\delta}}{\eta -\widetilde{\Delta}-\widetilde{\delta}}\cos \xi ,-\sin \xi,\cos \xi \right] ^{T},
\eea\ese
where the complex-valued angle $\xi $ is defined as
\be
\tan 2\xi =\frac{\widetilde{\Gamma} }{\widetilde{\Delta}-\widetilde{\delta}}.
\ee

Because the pulses are applied in the sequence ionizing-control-pump and the population is initially in state $\psi_{1}$,
 the initial adiabatic-state amplitudes are $b_{0}(-\infty )=1$, $b_{\pm }(-\infty )=0$. 
In the adiabatic limit the population $P_0(t)=\vert b_0(t)\vert^2$ of state $\adbstate _{0}(t)$ evolves as
\be
P_0(t)=\left\vert\exp \left[ -i\widetilde{\delta}(t)\int_{-\infty}^{t}
 \frac{\Omega^2(t^\prime)}{\widetilde{\Gamma}^2(t^\prime)-4\widetilde{\Delta}(t^\prime)\widetilde{\delta}(t^\prime)}dt^{\prime}\right] \right\vert^2.
\ee
The populations of the original states are 
\bse\label{Populations}
\bea
P_1(t) &=& \left\vert 1- \frac{\Omega^2(t)[\widetilde{\Gamma}^2(t)+4\widetilde{\delta}^2(t)]}
 {2[\widetilde{\Gamma}^2(t)-4\widetilde{\Delta}(t)\widetilde{\delta}(t)]^2}\right\vert^2 P_0(t), \label{P1} \\
P_{2}(t) &=& \left\vert \frac{2\Omega(t)\widetilde{\delta}(t)}{\widetilde{\Gamma}^2(t)-4\widetilde{\Delta}(t)\widetilde{\delta}(t)}\right\vert^2 P_0(t), \label{P2} \\
P_{c}(t) &=& \left\vert \frac{\Omega (t)\widetilde{\Gamma}(t)}{\widetilde{\Gamma}^2(t)-4\widetilde{\Delta}(t)\widetilde{\delta}(t)}\right\vert^2 P_0(t). \label{P3}
\eea
\ese
The fluorescence signal is calculated from Eq. \eqref{fluorescence} and \eqref{P2}, and then the ionization signal from Eq. \eqref{ionization3}.

In the following section we will use these formulas to examine how to minimize the fluorescence $F$, i.e. $P_{2}(t)$,
 and simultaneously to maximize the ionization $I$.

\section{Numerical Examples\label{Sec:Examples}}

We compare the analytical results derived in the preceding section with numerical simulations
 for the fluorescence signal of Eq. \eqref{fluorescence}, the ionization signal of Eq. \eqref{ionization3}, and the populations \eqref{Populations} of states $\state{1}$, $\state{2}$ and $\state{c}$,
 derived from numerical integration of Eq. \eqref{Schroedinger}, with the Hamiltonian \eqref{Hamiltonian}.
We assume Gaussian pulse shapes,
\bse\bea\label{Gaussians}
\Omega(t) &=& \Omega_0 e^{-\left( t-\tau \right)^2/T^2}, \\
\Gamma_i(t) &=& \Gamma_{i0} e^{-\left( t-\tau_i\right)^2/T_i^2}, \\
\Gamma_c(t) &=& \Gamma_{c0} e^{-\left( t-\tau_c\right)^2/T_c^2}.
\eea\ese
We use the pump pulse duration $T$ as a time unit and $1/T$ as a frequency unit, and choose the center of the pump pulse to define the zero reference point of time, $\tau=0$. 
All remaining parameters are variable: the peak pump Rabi frequency $\Omega_0$, the peak ionization rates $\Gamma_{i0}$ and $\Gamma_{c0}$,
 the centers of the ionizing and control pulses $\tau_i$ and $\tau_c$, their widths $T_i$ and $T_c$, the detunings $\Delta$ and $\delta$.
The Stark shifts are assumed zero because, as we have verified, they do not affect significantly the ionization signal.
For the Fano parameter we have chosen three values: $q=1$, $q=3$, $q=6$, which are close to the experimental values
 for LICS in sodium ($q=3.7$) \cite{sodium}, helium ($q=0.73$) \cite{helium}, and hydrogen atoms ($q=-5.9$) \cite{hydrogen},
 for electric-field mixing in rubidium ($q=3.3$) \cite{Feneuille},
 and for configuration mixing in potassium ($q=1$) \cite{Koide02a}, rubidium ($q=0.1-0.3$) \cite{Koide02b}, and cesium ($q \approx 0.43$) \cite{Koide02b}.

\begin{figure}[tb]
\centering\includegraphics[width=80mm]{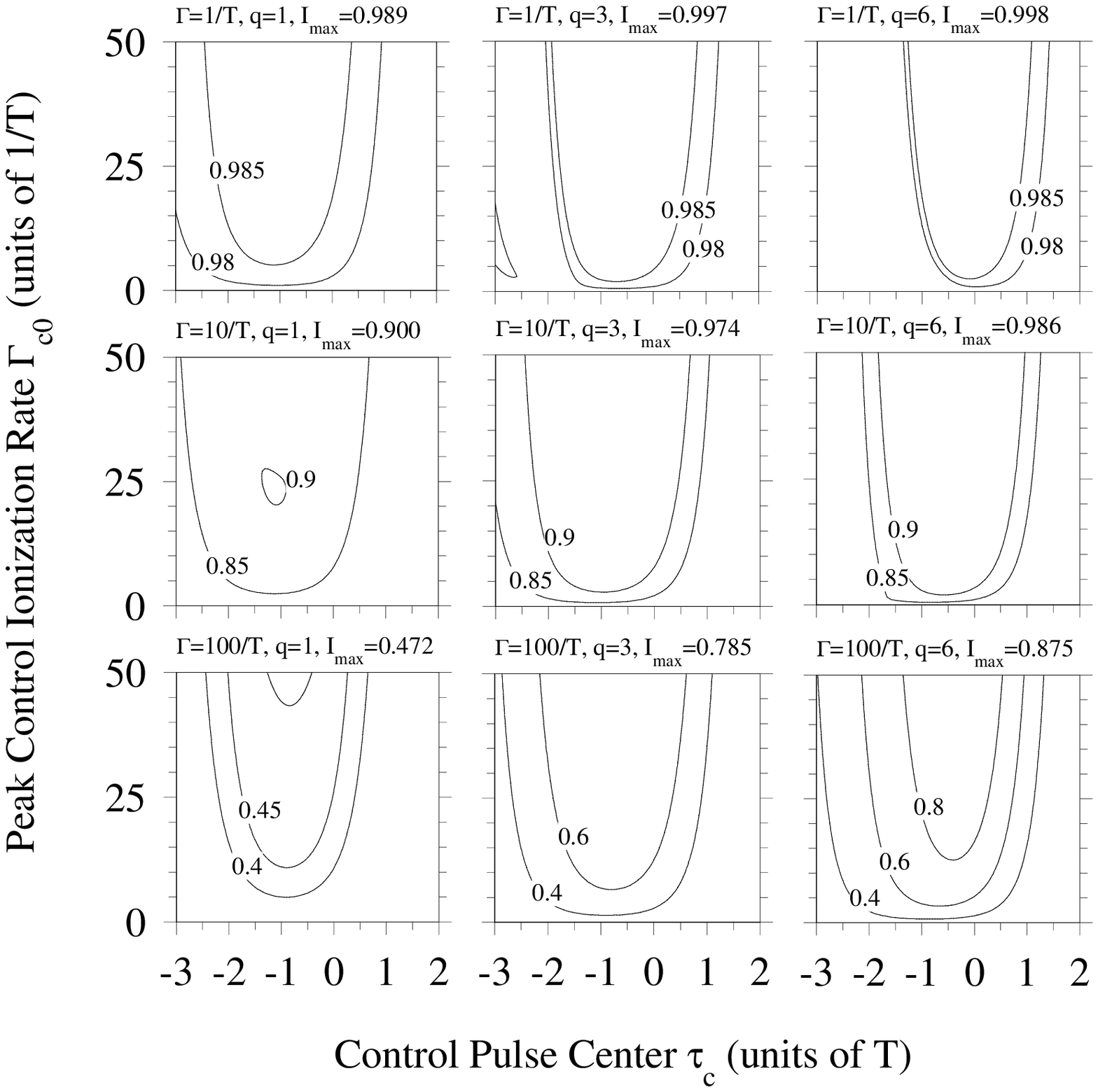}
\caption{Contour plots of the ionization signal as a function of the control pulse center $\tau_c$ and the  control ionization rate $\Gamma_{c0}$. 
The $\state{2}$ loss rate  is $\Gamma =1/T$ (top frames), $\Gamma=10/T$ (middle frames), or $\Gamma =100/T$ (bottom frames).
The Fano parameter is $q=1$ (left frames), $q=3$ (middle frames), or $q=6$ (right frames). 
The other parameters are $T_i=T_c=T$, $\Gamma_{i0}=50/T$, $\Omega_0=50/T$, $\tau_i=-T$, $\delta =10/T$, $\Delta =0$.
The number $I_{\rm max}$ atop each frame indicates the respective maximal ionization signal.
Without the control field  $I_{\rm max}$ is 0.978, 0.822, and 0.317 for $\Gamma=1$, 10, and 100, respectively.}
\label{Fig:tau-gamma}
\end{figure}

\begin{figure}[tb]
\centering\includegraphics[width=80mm]{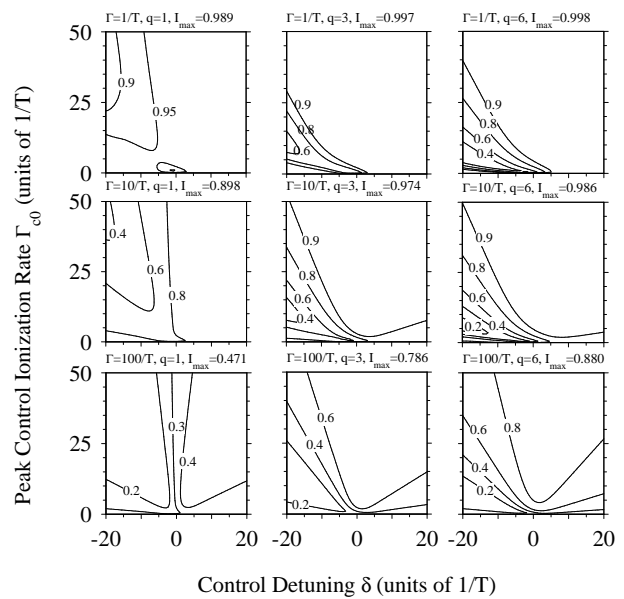}
\caption{Contour plots of the ionization signal as a function of the control laser detuning $\delta$ and the peak control ionization rate $\Gamma_{c0}$. 
The $\state{2}$ loss rate is $\Gamma =1/T$ (top frames), $\Gamma=10/T$ (middle frames), or $\Gamma =100/T$ (bottom frames).
The Fano parameter is $q=1$ (left frames), $q=3$ (middle frames), or $q=6$ (right frames). 
The other parameters are $T_i=T_c=T$, $\Gamma_{i0}=50/T$, $\Omega_0=50/T$, $\tau_i=-T$, $\tau_c=-0.5T$, $\Delta =0$.
The number $I_{\rm max}$ atop each frame indicates the respective maximal ionization signal.
Without the control field $I_{\rm max}$ is 0.978, 0.822, and 0.317 for $\Gamma=1$, 10, and 100, respectively.}
\label{Fig:delta-gamma}
\end{figure}

Figure \ref{Fig:tau-gamma} shows contour plots of the ionization signal as a function of the center of the control pulse $\tau_c$ and the peak control ionization rate $\Gamma _{c0}$
 for different values of the Fano parameter $q$ and the irreversible loss rate $\Gamma$ from state $\state{2}$.
Larger Fano parameters are clearly favourable for ionization but improvement is seen for $q=1$ too.
In principle a lower Fano parameter could be compensated by stronger ionizing and control fields.
For $\Gamma_{c0}=0$ (near the horizontal axis in each frame), the plain STIRAP into continuum, discussed in Sec. \ref{Sec:c-STIRAP}, occurs.
The bottom frames demonstrate that for strong loss ($\Gamma T\gg 1$) the LICS-STIRAP improves ionization dramatically, for example, from 0.317 without the control field to 0.875 with it.
We have verified that for larger laser intensities, a nearly complete ionization can be achieved.

Figure \ref{Fig:delta-gamma} shows contour plots of the ionization signal as a function of the control laser detuning $\delta $ and the control peak ionization rate $\Gamma_{c0}$.
Again, the presence of the control laser pulse, and the ensuing LICS, are essential in achieving high ionization signal, even for strong loss rate from state $\state{2}$. 
The asymmetry of the ionization signal vs the detuning is typical for the Fano LICS profile.

\begin{figure}[tbph]
\centering\includegraphics[width=80mm]{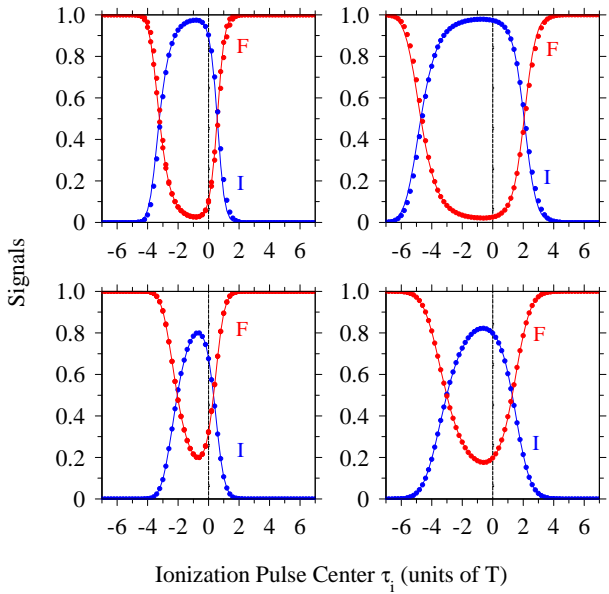}
\caption{(Color online) Ionization $I$ and florescence $F$ signals vs the center $\tau_i$ of the ionizing laser pulse. 
The $\state{2}$ loss rate is $\Gamma =10/T$ (top frames) or $\Gamma =100/T$ (bottom frames).
The pulse widths are $T_i=T_c=T$ (left frames) or $T_i=T_c=3T$ (right frames). 
The other parameters are $q=3$, $\Gamma_{i0}=\Gamma_{c0}=50/T$, $\Omega_0=50/T$, $\tau_c=-0.5T$, $\Delta=\delta =10/T$.
The solid curves show the analytical results derived from Eqs. \eqref{fluorescence}, \eqref{ionization3}, and \eqref{Populations}, the dots show the numerical results. }
\label{Fig:tau_i}
\end{figure}

In Fig. \ref{Fig:tau_i}, we display the fluorescence and ionization signals vs the timing of the ionizing pulse, for different values of the irreversible loss rate $\Gamma$
 and different widths of the control and ionizing pulses. 
The figure demonstrates that efficient ionization requires a counterintuitive pulse ordering, with the ionizing laser applied before the pump laser ($\tau_i<0$, with optimum about $\tau_i=-T$).
It also shows that an increase of the pulse widths (right frames) leads to broadening of the ionization profile but does not affect  appreciably the maximal ionization signal.
 The figure evidences an excellent agreement between the analytical theory and the numerical simulations.
This agreement indicates that indeed, the ionization dynamics occurs during the rising edge of the pump pulse, when the ionizing and control pulses are already present.
The manner in which the pulses terminate is not important, as evident when comparing the left frames (where the pump pulse starts and terminates later)
 and the right frames (where the pump pulse start later but terminates earlier because its width is shorter than the others).
This is one of the main differences between conventional STIRAP between discrete levels \cite{STIRAP} and LICS-STIRAP proposed here. 
In conventional STIRAP both the initial and final times of the pulses are important (the pump must start and terminate last). 
In LICS-STIRAP the initial times are important, but the final times are not, because there is no population left in the discrete states.
In this respect, LICS-STIRAP is similar to STIRAP between lossy states \cite{steering}.

\begin{figure}[tbph]
\centering\includegraphics[width=65mm]{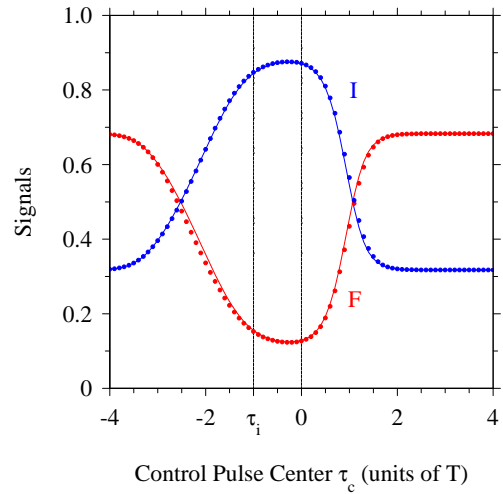}
\caption{(Color online) The ionization and florescence signals vs the center $\tau_c$ of the control laser pulse. 
The loss rate from state $\state{2}$ is $\Gamma =100/T$ and the pulse widths are $T_i=T_c=T$. 
The other parameters are $q=3$, $\Gamma_{i0}=\Gamma_{c0}=50/T$, $\Omega_0=50/T$, $\tau_i=-T$, $\Delta=\delta =10/T$.
The solid curves show the analytical results derived from Eqs. \eqref{fluorescence}, \eqref{ionization3}, and \eqref{Populations}, the dots indicate the numerical results. }
\label{Fig:tau_c}
\end{figure}

Figure \ref{Fig:tau_c} shows the fluorescence and ionization signals vs the timing of the control pulse. 
The figure demonstrates that the best timing between the ionizing and pump pulses is  at $\tau_c \sim \tau_i/2$, as used in other figures.
However, the technique is relatively robust against the control timing and moderate deviations from this prescription do not affect the ionization efficiency very much.
The background signal appearing for large deviations of $\tau_c$ from this region is produced by c-STIRAP. 
It  serves as a reference for the influence of LICS on STIRAP, significant in this case, once again.
This figure reveals another excellent agreement between analytical theory and numerical simulations.

\begin{figure}[tb]
\centering\includegraphics[width=65mm]{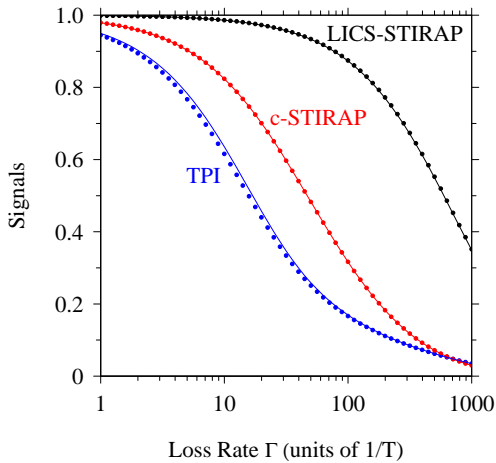}
\caption{(Color online) Ionization signal vs the $\state{2}$ irreversible loss rate $\Gamma$ for three different techniques. 
TPI: coincident pulses ($\tau_i=\tau=0$), no control pulse ($\Gamma_{c0}=0$);
 c-STIRAP: counterintuitively delayed pulses ($\tau_i=-T$, $\tau=0$), no control pulse ($\Gamma_{c0}=0$);
 LICS-STIRAP: counterintuitively delayed pulses ($\tau_i=-T$, $\tau=0$), with control pulse ($\Gamma_{c0}=50/T$, $\tau_c=-0.5T$).
The other parameters are $T_i=T_c=T$, $q=3$, $\Gamma_{i0}=50/T$, $\Omega_0=50/T$, $\Delta=\delta =10/T$.
The solid curves show the analytical results derived from Eqs. \eqref{fluorescence}, \eqref{ionization3}, and \eqref{Populations}. 
The dots display numerical results. }
\label{Fig:Gamma}
\end{figure}

Figure \ref{Fig:Gamma} shows the ionization signal as a function of the irreversible loss rate $\Gamma$ from state $\state{2}$.
The three curves show how the ionization efficiency decreases with $\Gamma$ for TPI, c-STIRAP (no control laser), and LICS-STIRAP.
The c-STIRAP efficiency is clearly superior to TPI, and LICS-STIRAP adds  further considerable improvement over c-STIRAP.
Note the horizontal logarithmic scale in $\Gamma$.
Remarkably, LICS-STIRAP maintains high ionization efficiency even when the intermediate state $\state{2}$ can decay hundreds of times during the interaction.
Again, the figure reveals an excellent agreement between the analytical theory and the numerical simulations.

\begin{figure}[tb]
\centering\includegraphics[width=65mm]{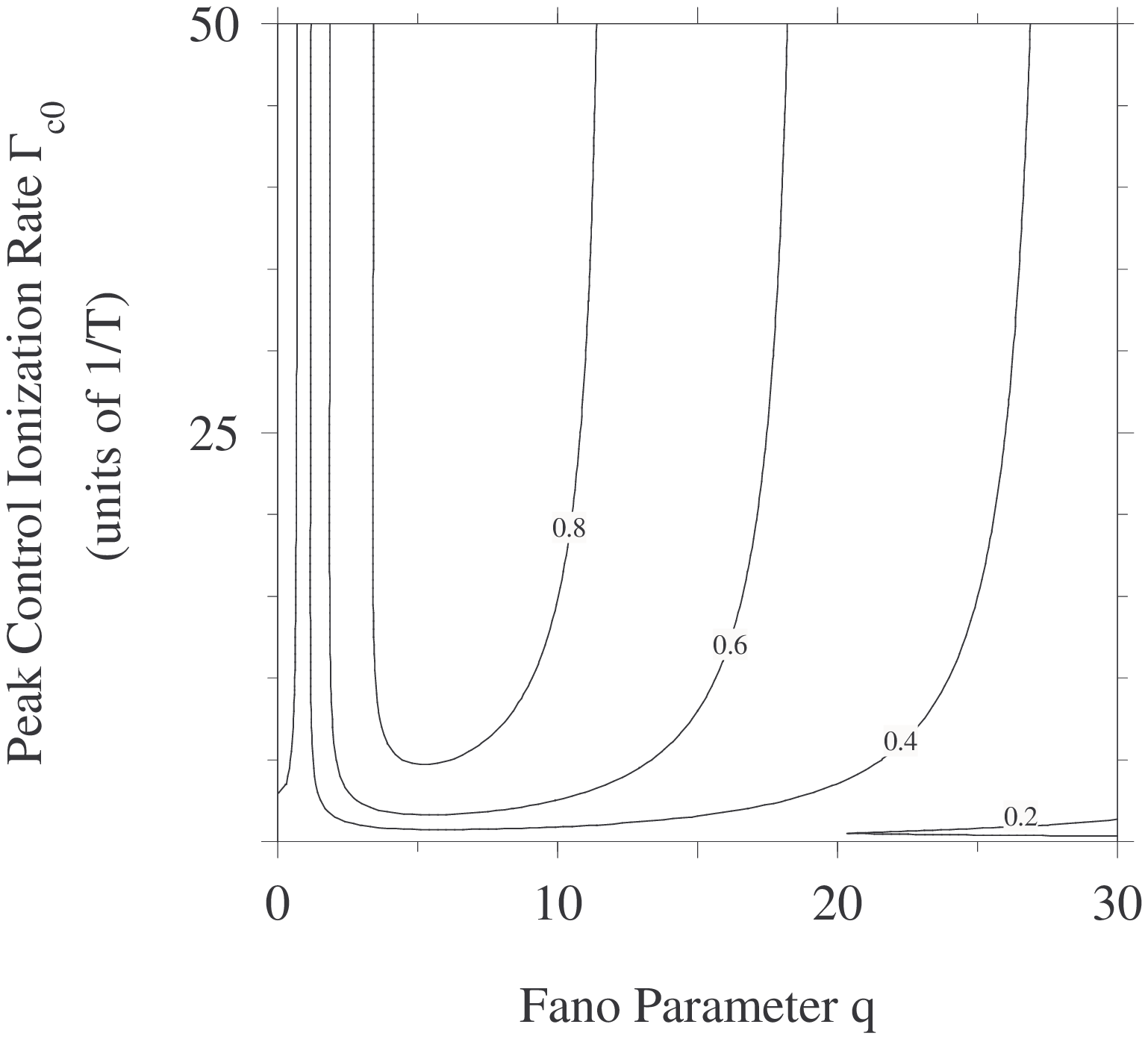}
\caption{Contour plot of the ionization signal as a function of the Fano parameter $q$ and the peak control ionization rate $\Gamma_{c0}$. 
The other parameters are $T_i=T_c=T$, $\Gamma_{i0}=50/T$, $\Omega_0=50/T$, $\Gamma =100/T$, $\tau_i=-T$, $\tau_c=-0.5T$, $\delta =10/T$,\ $\Delta =0$.}
\label{Fig:Fano}
\end{figure}

Figure \ref{Fig:Fano} shows the ionization signal vs the Fano parameter $q$ and the peak control ionization rate.
For moderate values of $q$ ($1\lesssim q \lesssim 10$) nearly complete ionization is achieved for sufficiently strong control pulses.
When $q$ is too small LICS is not sufficiently strong to simulate the presence of a  bound state and to create a quasi-dark state.
When $q$ is too large, a very large atomic coherence is created through the continuum, so that state $\state{c}$ is directly involved in the dynamics, with some atomic  population transfered to this state.

\section{Conclusions}\label{Sec:Conclusions}

We have demonstrated, using analytical techniques and numerical simulations, that STIRAP can be used as a tool for efficient ionization
 of the population of a discrete state $\state{1}$ coupled to a continuum via a lossy state $\state{2}$.
The ionizing laser must precede the pump laser coupling states $\state{1}$ and $\state{2}$, and both lasers must be strong enough to enforce adiabatic evolution.
The ionization probability is further enhanced when a LICS is created into the continuum and STIRAP is directed into this LICS
 because then a quasi-dark state composed of state $\state{1}$ and the LICS is created, via which the population flows into the continuum.
We have shown that almost complete ionization can be achieved even when the lifetime of the resonantly coupled state $\state{2}$ is much shorter than the laser interaction duration.
We have also shown that in the adiabatic limit, the main population dynamics takes place during the rising edge of the pump pulse, when it is lost irreversibly either via ionization or fluorescence.
Hence in LICS-STIRAP only the order in which the pulses arrive is important, a result significantly different from conventional STIRAP between discrete levels,
 where the order of the pulse terminations is also very important.

Experimental verification of these results will open opportunities for many applications, such as photoionization of ultracold atoms with efficiency close to unity and negligible population into intermediate discrete states. 
It should be noticed that LICS-STRAP relies on atomic structures in the continuum reached by appropriate laser sources.
An accurate examination of the atoms listed above and their experimental investigation shows that sodium is particularly suitable for experimental verification of the proposed technique.


\acknowledgments


This work has been supported by the EU ToK project CAMEL (Grant No. MTKD-CT-2004-014427), the EU RTN project EMALI (Grant No. MRTN-CT-2006-035369),
 and the Bulgarian National Science Fund grant WU-205/06.
AAR thanks the Department of Physics at Pisa University for the hospitality during his visit there. 
The authors acknowledge useful discussions with L.P. Yatsenko and P. Lambropoulos.


\end{document}